\documentclass[aps,prl,reprint,groupedaddress]{revtex4-1}

\usepackage{graphicx}  
\usepackage{dcolumn}   
\usepackage{bm}        
\usepackage{amssymb}   

\begin{document}

\title{Stochastic modeling of regulation of gene expression by multiple
small RNAs}

\author{Charles Baker}
\email{cbaker@vt.edu}
\author{Tao Jia}
\email{tjia@vt.edu}
\author{Rahul V. Kulkarni}
\email{kulkarni@vt.edu}
\affiliation{Department of Physics, \\
Virginia Polytechnic Institute and State University, \\
Blacksburg, VA 24061}
\date{\today}

\begin{abstract}

A wealth of new research has highlighted the critical roles of
small RNAs (sRNAs) in diverse processes such as quorum sensing
and cellular responses to stress. The pathways controlling these
processes often have a central motif comprising of a master regulator protein
whose expression is controlled by multiple sRNAs. However, the regulation of
stochastic gene expression of a single target gene by multiple
sRNAs is currently not well understood. To address this issue, we
analyze a stochastic model of regulation of gene expression by
multiple sRNAs. For this model, we derive exact analytic results
for the regulated protein distribution including compact
expressions for its mean and variance. The derived results
provide novel insights into the roles of multiple sRNAs in
fine-tuning the noise in gene expression. In particular, we show
that, in contrast to regulation by a single sRNA, multiple sRNAs
provide a mechanism for independently controlling the mean and
variance of the regulated protein distribution.
\end{abstract}

\maketitle


Small non-coding RNAs (sRNAs) are known to play a central role in
diverse cellular pathways that bring about global changes in gene
expression \cite{Waters2009}.  In several cases, such global changes
are coordinated by a master regulator protein whose expression
is controlled by multiple sRNAs \cite{Beisel2010,Peter2010}. Examples include 
regulation of the master regulator in bacterial quorum-sensing
pathways by multiple sRNAs \cite{Lenz2004} and regulation of
the alternative sigma factor $\sigma^{s}$ by four distinct sRNAs, each of which responds
to different environmental stresses \cite{Beisel2010}.  Despite its
importance in coordinating cellular stress responses and related
processes, the role of multiple sRNAs in regulating the expression of a
single target gene is not yet well understood \cite{Peter2010}. In
this work, we address this issue by analyzing a
stochastic model that elucidates potential functional roles for
this widely-occurring regulatory motif.

Regulation of gene expression by sRNAs is a post-transcriptional
process: sRNAs can bind to messenger RNAs (mRNAs) and control protein
production by altering mRNA stability or by
regulating translational efficiency \cite{Waters2009}.  The intrinsic
stochasticity of the underlying biochemical reactions can produce
significant variations (`noise') in gene expression among individual cells in isogenic populations
\cite{Kaern2005,Paulsson2005,Maheshri2007,Raj2008,Kondev2008}.
Although noise in gene expression can
have deleterious effects in some cases and thus needs to be limited; in other cases such noise is utilized and
indeed required by the cell e.g.\ for processes leading to cell-fate
determination \cite{Losick2008,Hornstein2006}.  Furthermore, it has
been argued that noise in gene expression could be advantageous under
conditions of high stress, since variability in a population provides
a bet-hedging strategy that can enable survival
\cite{Fraser2009,Eldar2010}. Regulation of the noise in gene expression
is thus essential for the proper functioning of several cellular processes.
Since sRNAs regulate critical
cellular processes, understanding their role in fine-tuning the noise
in gene expression is of fundamental importance
\cite{Hornstein2006}.

A quantitative understanding of the cellular functions of sRNAs is
aided by the development of models, which can often produce insights
that guide future experiments.  In recent research, several models for
regulation by sRNAs have been developed.  Since many sRNAs are known
to repress gene expression, most previous models have focused on
regulation by irreversible stoichiometric degradation
\cite{Levine2007,Mitarai2007,Mehta2008,Frohlich2009}. However, sRNAs
can affect not only mRNA degradation rates but also protein production
rates and the corresponding biochemical reactions are, in general,
reversible \cite{Flynt2008,Waters2009}. Furthermore, not all sRNAs
repress gene expression; there are sRNAs which are known to activate
gene expression and even some which can switch from activating to
repressing in response to cellular signals
\cite{Flynt2008,Vasudevan2007}.  To quantify the corresponding effects
on stochastic gene expression, a general model which includes the
different mechanisms of sRNA-based regulation needs to be analyzed.
Such a model, for the case of a single sRNA regulator, has been
developed in recent work \cite{Jia2010}.  Analysis of this model and
its extension to multiple sRNAs thus provides a means of addressing
outstanding questions about the impact of different modes of
regulation by sRNAs on the noise in gene expression.

In this work, we generalize our previous model \cite{Jia2010} to
analyze the case of multiple sRNAs regulating a single mRNA
target. Specifically, we derive exact analytic expressions for the
generating function of a protein burst distribution resulting from the
regulation of a single target by an arbitrary number of sRNAs.  Using
this expression, we obtain compact analytic expressions for both the
mean and variance of the regulated protein distribution. We first analyze
these results for the case of a single regulator and derive
insights into different modes of regulation by sRNAs. These results
are then contrasted with features unique to regulation by multiple
sRNAs. In particular, we show that, in contrast to regulation by a
single sRNA, regulation by multiple sRNAs provides the cell with a
mechanism to independently fine-tune both the mean and variance of the
regulated protein distribution.

\begin{figure}[tb]
\resizebox{8cm}{!}{\includegraphics{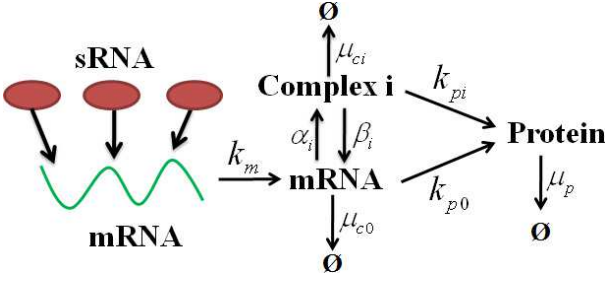}}
\caption{Schematic illustration of regulation of gene expression by
multiple sRNAs. In the full reaction scheme, there are $N$ different
regulators and the kinetic scheme is shown for the $i^{\mathrm{th}}$
sRNA regulator. The association and dissociation rates for binding to
the mRNA are denoted by $\alpha_{i}$ and $\beta_{i}$
respectively. Association results in a complex which produces proteins
with rate $k_{p_{i}}$ and is degraded with rate $\mu_{c_{i}}$. Note
that for the mRNA to transition from one complex to
another, it must first return to its unbound state before forming
a new complex.  \label{fig1}}
\end{figure}\noindent

We begin by considering protein production from a single mRNA
regulated by $N$ independent sRNAs. The corresponding reaction scheme
is shown in Figure 1.  The mRNA has $N+1$ possible states, with the
states $i=1,..,N$ denoting mRNA bound to the $i^{\mathrm{th}}$
sRNA regulator to form complex $i$.  For notational simplicity, we
denote the unbound mRNA state as complex $0$. An unbound mRNA forms
complex $i$ with rate $\alpha_{i}$ and the complex can either
dissociate with a rate $\beta_{i}$, decay with a rate $\mu_{c_{i}}$,
or initiate protein production with a rate $k_{p_{i}}$. We assume the
sRNA regulators are present in large amounts such that fluctuations in
their concentration can be ignored; correspondingly the rates
$\alpha_{i}$ are taken to be constant.

The distribution of proteins produced from a single mRNA (interacting
with $N$ sRNAs) before it decays is denoted as the protein burst
distribution $P_{b,N}(n)$. We further define the function $f_{i}(n,t)$
which denotes the probability that $n$ proteins have been produced and
the mRNA is in state $i$ at time $t$.
Correspondingly, the protein burst distribution,
$P_{b,N}(n)$, can be obtained from
\begin{equation}
P_{b,N}(n)=\sum_{i=0}^{N}\intop_{0}^{\infty}f_{i}(n,t)\mu_{c_{i}}dt.
\end{equation}
The time-evolution of the probabilities $f_{i}(n,t)$ is governed by
the Master equation
\begin{eqnarray}
\frac{\partial f_{0}(n,t)}{\partial t} & = &
 k_{p_{0}}(f_{0}(n-1,t)-f_{0}(n,t))\nonumber \\ & &
 -(\mu_{c_{0}}+\sum_{i=1}^{N}\alpha_{i})f_{0}(n,t)+\sum_{i=1}^{N}\beta_{i}f_{i}(n,t)\nonumber
 \\ \frac{\partial f_{i}(n,t)}{\partial t} & = &
 k_{p_{i}}(f_{i}(n-1,t)-f_{i}(n,t))\nonumber \\ & &
 -(\mu_{c_{i}}+\beta_{i})f_{i}(n,t)+\alpha_{i}f_{0}(n,t)
\end{eqnarray}
The initial condition
corresponds to creation of a single unbound mRNA and no proteins in the
system at time $t=0$, i.e. $f_{0}(0,0)=1$.
The procedure outlined in \cite{Jia2010} can now be applied to obtain
the burst distribution or more precisely the corresponding generating
function $G_{b,N}(z) = \sum_{n} z^{n} P_{b,N}(n)$.
Specifically, defining
$F_{i}(z,s)=\sum_{n}{z^n\int_{0}^{\infty}{e^{-st}f_{i}(n,t)}\,dt}$, we
obtain the generating function using
\begin{equation}
G_b(z) = \lim _{s\to 0} \sum_{i=0}^{N} \Bigl(\mu_{c_{i}} F_{i}(z,s) \Bigr)\label{eq:Gb_single}
\end{equation}
To present the results in a compact form, it is convenient to define the
dimensionless variables  $\xi_{i} =
\frac{k_{p_{i}}}{\beta_{i}+\mu_{c_{i}}}$ and $\omega_{i}
= \frac{\alpha_{i}}{\beta_{i} + \mu_{c_{i}}} \left(\frac{\mu_{c_{i}}}{\mu_{c_{0}}}\right)$ for $i>0$
and $n_{i} = \frac{k_{p_{i}}}{\mu_{c_{i}}}$ for $i \ge 0$.  Now, by
setting $\omega_{0}=1$ and $\xi_{0}=0$ we further define
the `weight functions'
$\omega_{i}(z) =\omega_{i}\frac{1}{1 + \xi_{i}(1-z)} \label{eq:omega}$.
Note that $\frac{1}{1 + \xi_{i}(1-z)}$ is the generating function of a
geometric distribution with mean $\xi_{i}$.

Using the above definitions, we obtain the following exact expression
for the generating function of the protein burst distribution
\begin{equation}
G_{b,N}(z)=\frac{\sum_{i=0}^{N}\omega_{i}(z)}{\sum_{i=0}^{N}\omega_{i}(z) + \sum_{i=0}^{N}\omega_{i}(z)n_{i}(1-z)}\label{eq:generating_func}
\end{equation}
For $N=0$, i.e.\ the unregulated case, the generating function reduces to
\begin{equation}
G_{b,0}(z) = \frac{1}{1 + n_{0}(1-z)} \label{eq:unregulated}
\end{equation}
in agreement with previous work showing that
the protein burst
distribution is a geometric distribution with mean
$n_{0} = \frac{k_{p_{0}}}{\mu_{c_{0}}}$ \cite{Berg1978,Jia2010}.
Eq. (\ref{eq:generating_func}) provides the generalization of this
result for the case of $N$ sRNA regulators using the  weight
functions $\omega_{i}(z)$.

An important mechanism of regulation by sRNAs corresponds to the case that
sRNA binding prevents ribosome access and thus blocks translation.
For the case that all the regulators act to
fully repress translation, i.e.\ $k_{p_{i}}=0$ for $i>0$, we
have $\omega_{i}(z)=\omega_{i}$. Correspondingly, the generating
function reduces to the formula for the unregulated
case (Eq. (\ref{eq:unregulated})) with a renormalized mean given by
$\frac{n_0}{\sum_{i=0}^{N}\omega_{i}}$.
This interesting observation
indicates that regulation by sRNAs which function by fully repressing
translation is reversible: for arbitrary concentrations of the sRNA regulator,
by appropriately adjusting the parameter $k_{p_{0}}$
(e.g.\ by adjusting the concentration of ribosomes),
the regulated protein distribution in the presence of sRNAs
can be made identical to the distribution for the unregulated case
(prior to introduction of the sRNAs).

For the general case, the generating function can be recast in a form
that shows that the protein burst distribution is a mixture of $N+1$
geometric distributions \cite{Jia2010}. However, the corresponding
expression, even for the case of $N=2$, is too complex to be
reproduced here.  On the other hand, using
Eq. (\ref{eq:generating_func}), compact analytic expressions for the
mean, $n_{b_{N}}$, and squared coefficient of variance,
$\sigma_{b_{N}}^{2}/n_{b_{N}}^{2}$ can be derived.
The mean (scaled by the unregulated mean) is
given by
\begin{equation}
\frac{n_{b_{N}}}{n_{b_{0}}}=1+F_{N}\label{eq:N_burst_mean}
\end{equation}
 and the noise strength (squared coefficient of variance) is given by
\begin{equation}
\frac{\sigma_{b_{N}}{}^{2}}{n_{b_{N}}{}^{2}}=1+\frac{1}{n_{b_{N}}}+Q_{N}\label{eq:N_burst_noise}
\end{equation}
 where
\begin{eqnarray*}
F_{N} & = &
\frac{\sum_{i=0}^{N}\omega_{i}\left(n_{i} - n_{0}\right)}{\sum_{i=0}^{N}\omega_{i}n_{0}}\\
\\Q_{N} & = &
\frac{\sum_{i,j=0}^{N}\omega_{i}\omega_{j}(\xi_{i}-\xi_{j})(n_{i}-n_{j})}{\left(\sum_{i=0}^{N}\omega_{i}n_{i}\right)^{2}}\label{eq:QN} \\
\end{eqnarray*}
Note that the signs of $F_{N}$ and $Q_{N}$ characterize the impact of
the sRNAs on the regulated protein distribution. Specifically, the
unregulated case has mean $n_{b_{0}}$; thus $F_{N}<0$ corresponds to
repression whereas $F_{N}>0$ corresponds to activation. Similarly, an
unregulated protein burst distribution with mean $n_{b_{N}}$ has a
squared coefficient of variance $1+1/n_{b_{N}}$; thus, when $Q_{N}<0$
we have noise reduction whereas $Q_{N}>0$ corresponds to increased
noise strength (relative to an unregulated burst distribution with the same
mean).

We now focus on using Eq. (\ref{eq:N_burst_mean}) and Eq. (\ref{eq:N_burst_noise}), to elucidate interesting features for the
case of regulation by a single sRNA, i.e. $N=1$.
Note that all of the variables in the expressions for the mean and
noise strength are always positive (or zero) except for the term
$\left(n_{1} - n_{0}\right)$. Thus, the sign of $F_{1}$ and $Q_{1}$ is
determined completely by $\Delta_{10} = n_{1}- n_{0}$.
When $\Delta_{10}>0$ both the mean and the noise strength are higher than
their unregulated values.  Similarly, when $\Delta_{10}<0$ both the
mean and the noise strength are lower than the corresponding unregulated values
(except for the case $\xi_i =0$ for which the noise strength is identical to
an unregulated distribution with the same mean).
In either case, we note that, for a single sRNA regulator,
there is a coupling between the mean and
noise strength of the regulated burst distribution such that
both cannot be tuned independently, e.g.\ a decrease in the mean
cannot be associated with an increase in the noise strength.

\begin{figure}[tb]
\centering
\begin{center}
   $\begin{array}{c@{\hspace{0.05cm}}c}
   \multicolumn{1}{l}{\mbox{\bf (a)}} &
   \multicolumn{1}{l}{\mbox{\bf (b)}} \\ [-0.0cm]
       \resizebox{45mm}{!}{
       \includegraphics*{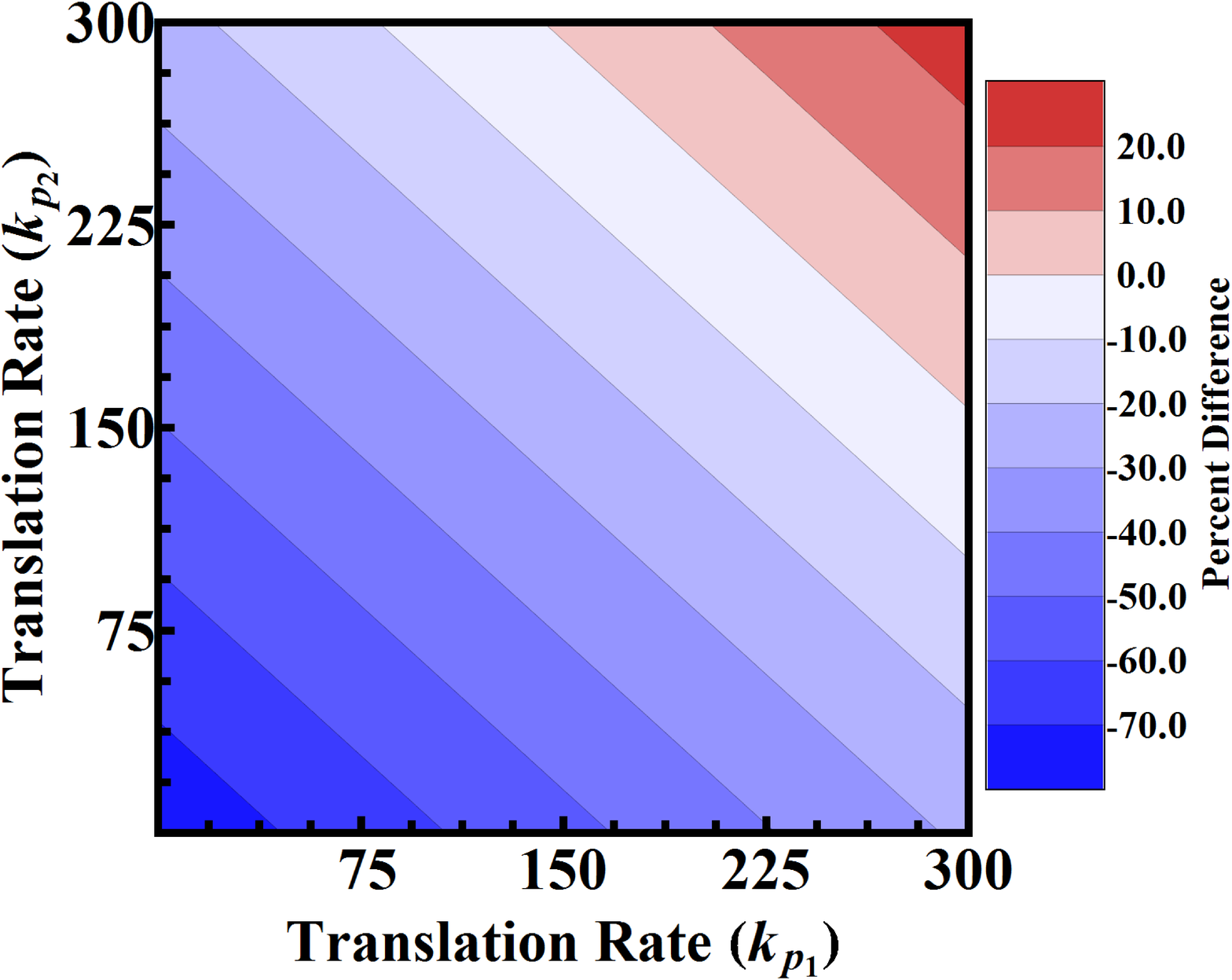}} &
       \resizebox{45mm}{!}{
       \includegraphics*{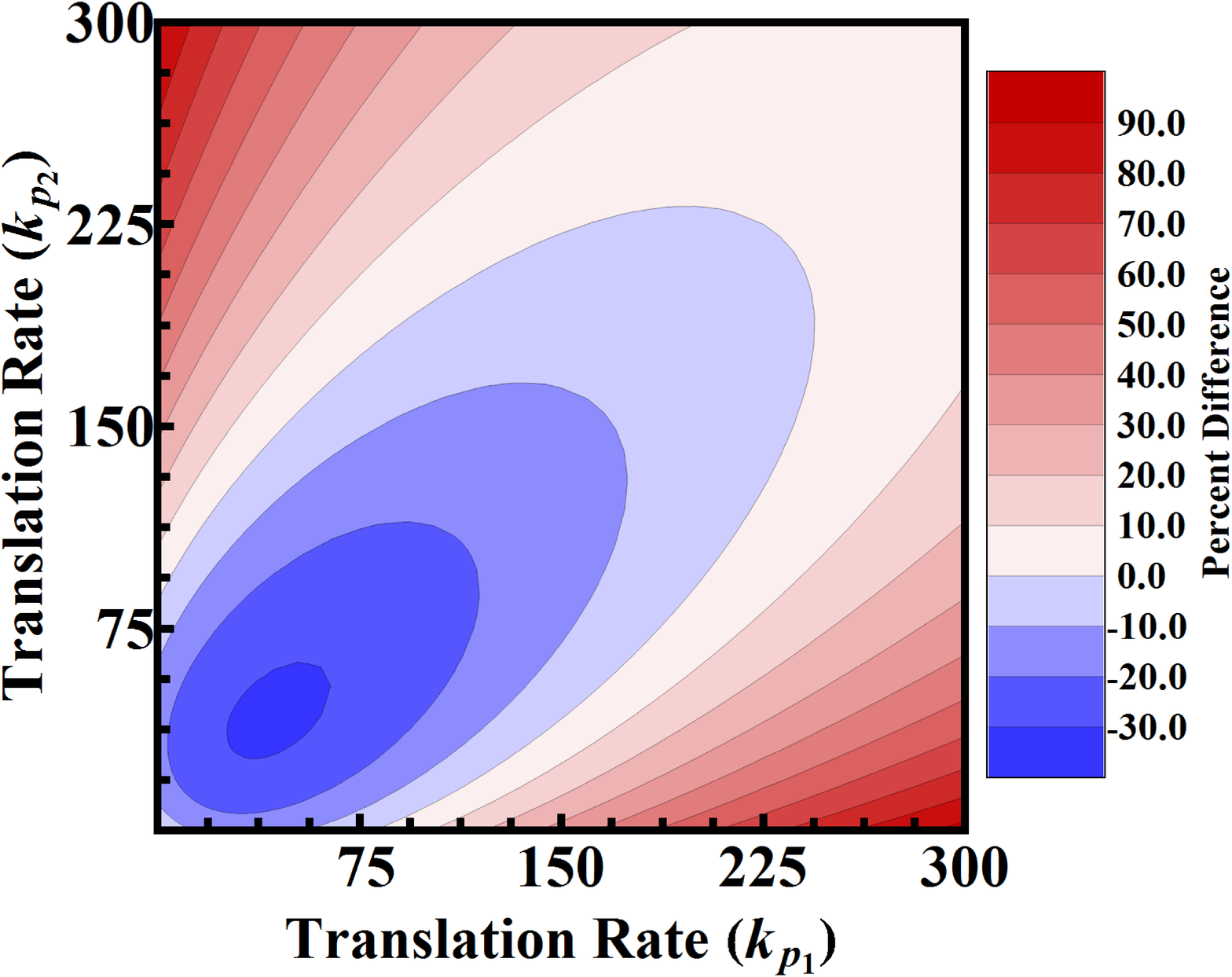}}\\
   \end{array}$
\end{center}
\caption{Contour plots for the percent change in the mean and noise strength of a two regulator system from the corresponding unregulated values as a function of $k_{p_{1}}$ and $k_{p_{2}}$. (a) Mean: Plot of $f(k_{p_{1}},k_{p_{2}})=\frac{n_{b_{2}}-n_{b_{0}}}{n_{b_{0}}}\cdot100\%$. Note
that along the contour $f(k_{p_{1}},k_{p_{2}})=-20\%$ the noise strength changes from less than $-5\%$ to over $70\%$. (b) Noise Strength: Plot of $g(k_{p_{1}},k_{p_{2}})=\frac{Q_{2}}{1+1/n_{b_{2}}}\cdot 100 \%$. Note
that $g(k_{p_{1}},k_{p_{2}})$ contains contours that sweep out a large
portion of the plotted $(k_{p_{1}},k_{p_{2}})$ state space. By proportionally changing the $k_{p}$ values corresponding to the two regulators, the noise strength can be varied while maintaining the same mean value. The parameters used were $k_{p_{0}}=50$, $\mu_{c_{0}}=1$, $\mu_{c_{1}}=4.5$,
$\mu_{c_{2}}=4.5$, $\beta_{1}=1$, $\beta_{2}=0.5$, $\alpha_{1}=2$ and
$\alpha_{2}=2$.
\label{fig2}}
\end{figure}

In contrast to the case of regulation by a single sRNA, in the case of
regulation by multiple sRNAs, the mean and noise of the protein
distribution can be tuned independently.  The deviation of the mean
from its unregulated value depends solely upon terms of the form
$\Delta_{i0} = n_{i} - n_{0}$.  On the other hand, considering the
general form of the noise strength for $N>1$, we have terms of the form
$\tilde{\Delta}_{ij}=(\xi_{i}-\xi_{j})(n_{i}-n_{j})$ that contribute
to the deviation from the corresponding unregulated value. Thus, for
appropriately chosen parameters, two sRNAs can be used to tune both the
mean and variance of the regulated protein distribution as discussed
below.

Consider the case of regulation by 2 sRNAs that are maintained at some
fixed cellular concentrations. A new mRNA target for these sRNAs can arise
from the evolution of appropriate sRNA binding sites on the mRNA
sequence. For the new target, we assume that the parameters $k_{p_{1}}$
and $k_{p_{2}}$ can be tuned based on changes in the sequence and location
of the sRNA binding sites. The corresponding variation in the mean and noise
strength is shown in Fig. 2. Note that by maintaining a linear
relationship between $k_{p_{1}}$ and $k_{p_{2}}$, the mean of the
regulated protein distribution can be left unchanged; however, the noise
strength can be tuned over a large range. For example, for some choices of
the parameters, the mean can be fixed and the noise strength can
be varied by over $100\%$ relative to the unregulated
distribution (see Fig. 2).
In this context, it is interesting to note that it has been observed that
several sRNAs have a minimal effect on the mean levels of their regulatory
targets.
For such targets, sRNAs
could be functioning primarily as modulators of noise while giving
rise to only a minimal change in mean levels due to regulation
\cite{Hornstein2006}. Our results provide quantitative insight into how
such regulation can be implemented using multiple sRNA regulators.

\begin{figure}[tb]
\resizebox{8cm}{!}{\includegraphics{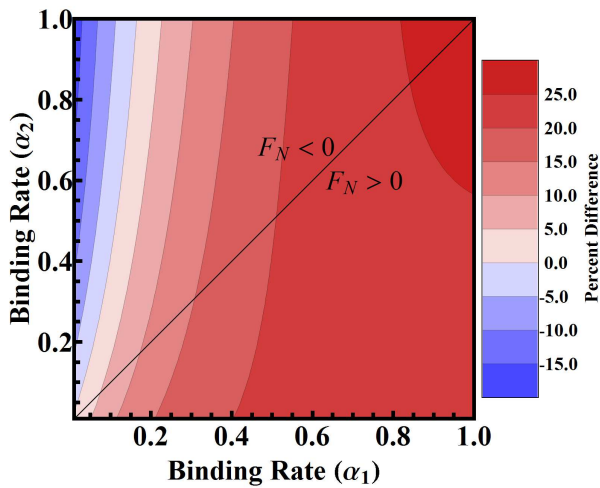}}
\caption{Contour plot of the percent change in noise strength of a two regulator pathway from its corresponding unregulated value as a function of $\alpha_{1}$ and $\alpha_{2}$, i.e.\ $f(\alpha_{1},\alpha_{2})=\frac{Q_{2}}{1+1/n_{b_{2}}}\cdot 100 \%$. The change in mean from the unregulated to
regulated pathways, $F_{N}$, is positive below and negative above the
line $\alpha_{1}=\alpha_{2}$. The parameters used were $k_{p_{0}}=50$,
$k_{p_{1}}=200$, $k_{p_{2}}=72.5$, $\mu_{c_{0}}=1$,
$\mu_{c_{1}}=2.725$, $\mu_{c_{1}}=2.725$, $\beta_{1}=0.15$ and
$\beta_{2}=0.15$. \label{fig3}}
\end{figure}\noindent

The results obtained also illustrate how changing sRNA concentrations
can be used to modulate the noise in gene expression.
For our model, changes in the
concentration of the sRNA regulators effectively alter the binding
rates ($\alpha_{i}$) to the mRNA. From Eq. (\ref{eq:N_burst_mean}), we
see that for 2 regulators, by choosing one of the regulators to be a
repressor and the other to be an activator, the mean of the regulated
protein distribution can be increased ($F_{N}>0$) or decreased ($F_{N}<0$)
by adjusting the relative concentrations of the two regulators.
Furthermore, by changing the concentrations of the regulators such
that their relative concentration is fixed, the mean of the
regulated protein distribution is left unchanged, whereas the variance
can be tuned over a range of values.
This insight is particularly
relevant, given that noise can be advantageous to a cell, especially in
response to stress.

Finally, we note that while the above analysis focuses on the burst
distribution from a single mRNA, the results are readily generalized
to the case of arbitrary mRNA burst distributions,
assuming independent contributions from
each mRNA \cite{Jia2011}. Assuming independent bursts and using
equations derived in recent work \cite{Jia2011}, we can use the
derived results to obtain corresponding expressions for the mean and
variance of the steady-state protein distribution. A detailed analysis
of the connection to steady-state distributions will be presented elsewhere.

Modulation of gene expression in pathways containing multiple
sRNA regulators could provide a multitude of selective
advantages in specific contexts and our model provides a means of
gaining insight into these processes. In particular, our analysis
shows how by adjusting the concentrations of multiple sRNA regulators,
a cell can initiate finely tuned
responses to external stimuli.  This could explain
the ubiquity of sRNA regulators in cellular stress response, a process
for which gene expression noise is known to be critical. In a broader
context, the fine control afforded to cells via regulation by multiple
sRNAs could be useful in signal integration when multiple environmental
stimuli are present simultaneously.
This work provides a basis for future work investigating
such signal integration and response in more complex pathways;
thus opening new avenues for understanding and modeling
stochastic gene expression in a wider class of regulatory networks. \\

\begin{acknowledgments}
The authors acknowledge funding support from the NSF through award PHY-0957430.
\end{acknowledgments}


\bibliographystyle{apsrev}
\bibliography{srna_1_30_2011}

\end{document}